\documentclass[11pt,fleqn,amstex]{article}

\usepackage{latexsym}

\usepackage{amsmath}
\usepackage{amsthm}
\usepackage{amssymb}
\usepackage{amsfonts}
\usepackage{color}

\newcommand{\rf}[1]{(\ref{#1})}

\newcommand{\ba}{\begin{array}}
\newcommand{\ea}{\end{array}}

\newcommand{\be}{\begin{equation}}
\newcommand{\ee}{\end{equation}}
\newcommand{\ods}{\par \vspace{0.2cm} \par}

\newcommand{\const}{{\rm const}}

\newcommand{\Cl}{{\cal C}}

\newcommand{\e}{{\bf e}}

\newcommand{\E}{{\mathbb E}}

\newcommand{\ep}{\varepsilon}

\newcommand{\no}{\noindent}

\newtheorem{prop}{Proposition}[section]
\newtheorem{Th}[prop]{Theorem}

\newtheorem{rem}[prop]{Remark}
\newtheorem{cor}[prop]{Corollary}

\begin{document}

\title{  Group interpretation of the spectral parameter. The case of isothermic surfaces}
\author{ Jan L.\ Cie\'sli\'nski\thanks{E-mail: \ 
 j.cieslinski@uwb.edu.pl},
\\ {\footnotesize Uniwersytet w Bia\l ymstoku, Wydzia{\l} Fizyki}
\\ {\footnotesize ul.\ Cio\l kowskiego 1L,  15-245 Bia\l ystok, Poland}
\\[2ex] {  Artur Kobus}
\\ {\footnotesize  Politechnika Bia\l ostocka, Wydzia{\l} Budownictwa i In\.zynierii \'Srodowiska}
\\ {\footnotesize  ul.\  Wiejska 45E,  15-351 Bia\l ystok, Poland}
}

\date{ }

\maketitle

\begin{abstract}
It is well known that in some cases the spectral parameter has a group interpretation. We discuss in detail the case of Gauss-Codazzi equations for isothermic surfaces immersed in $\E^3$. The algebra of Lie point symmetries is 4-dimensional and all these symmetries are also symmetries of the Gaus-Weingarten equations (which can be considered as $\mathfrak so(3)$-valued non-parametric linear problem). In order to obtain a non-removable spectral parameter one has to consider $\mathfrak so(4,1)$-valued linear problem which has a 3-dimensional algebra of Lie point symmetries. The missing symmetry introduces a non-removable parameter.  In the second part of the paper we extend these results on the case of isothermic immersions in arbitrary multidimensional Euclidean spaces. In order to simplify calculations the problem was formulated in terms of a Clifford algebra.    
\end{abstract}

\noindent {\it Mathematics Subject Classification 2010}: 58D19; 53C42; 37K10; 53A04

\par 
\noindent {\it Keywords}: isothermic surfaces, Lie point symmetries, non-removable  parameter, Lax pair, Gauss-Codazzi equations, Clifford algebra
\par

\section{Introduction}

Integrable systems of nonlinear partial differential equations often arise as compatibility conditions for an associated linear system, known as the Lax pair, with the so called spectral parameter  (see, for instance, \cite{BBT}). On the other hand, in the geometry nonlinear systems (Gauss-Codazzi equations) with associated linear problems (Gauss-Weingarten equations) are ubiquitous but the linear problem is usually without a parameter. Therefore it is natural to expect that one of symmetries of the nonlinear system is not a symmetry of the linear problem and can be used to produce the spectral parameter, provided that this parameter is non-removable (i.e., it cannot be removed by a gauge transformation). This idea has been suggested as a working criterion of integrability \cite{LS,LST}, soon formulated in a purely algebraic way \cite{Ci-isolate,CGS2}. Around the same time this the same idea has been formulated in terms of coverings \cite{KV}. Since then many authors used Lie symmetries to obtain or discuss Lax pairs with non-reomvable parameter, see  \cite{BKMV,Ci-nonl,Ci-group,ELS,Mo}. More powerful approach by using coverings has been developed in this context as well, see \cite{KM,Ma}.

The group interpretation of the spectral parameter is rather natural but still  surprisingly small number of cases is checked and confirmed. 
In the first part of this paper we discuss isothermic immersions in $\E^3$ \cite{CGS1,CG,Eis}. The corresponding Gauss-Codazzi equations have 4-dimensional algebra of Lie point symmetries. All these symmetries are symmetries of the  $\mathfrak so(3)$-valued linear problem (classical Gauss-Weingarten equations), as well. This is consistent with  non-existence of an $\mathfrak so(3)$-valued Lax pair for isothermic surfaces. In order to obtain the classical Lax pair for isothermic surfaces one has to consider a larger matrix algebra, namely $\mathfrak so(4,1)$. Then, one of the symmetries of Gauss-Codazzi equations is not a symmetry of the $\mathfrak so(4,1)$-valued linear problem and we recover the Lax pair with a non-removable spectral parameter. 
In the second part of the paper we extend these results on the case of isothermic surfaces immersed in multidimensional Euclidean spaces. Thus we successfully checked that yet another class of integrable systems admits the  group interpretation of  the spectral parameter.

\section{Lie point symmetries of Gauss-Codazzi equations for isothermic surfaces}

Isothermic surfaces (isothermic immersions in ${\mathbb E}^3$) are defined by the property that their curvature lines admit conformal (``isothermic'') parameterization \cite{CGS1,Eis}. In other words, there exists conformal cordinates such that both fundamental forms are diagonal:
\be  \label{forms}
I = e^{2 \theta} (du^2 + dv^2) ,  \\[1ex]
II = e^{2\theta} (k_1 du^2 + k_2 dv^2) \ ,
\ee
where $\theta = \theta (u, v)$ defines conformal coordinates while $k_1 = k_1 (u,v)$ and $k_2 = k_2(u,v)$ are principal curvatures. These functions have to satisfy the system of Gauss-Codazzi equations:
\be \ba{l}
\label{gmc}
\theta_{,uu} + \theta_{,vv} + k_1 k_2 e^{2 \theta} = 0, \\[1ex] 
 k_{1,u} + (k_1 - k_2) \theta_{,u} = 0, \\[1ex]
 k_{2,v} + (k_2 - k_1) \theta_{,v} = 0. 
\ea \ee
Lie point symmetries of \rf{gmc} can be computed in the standard way (see \cite{O}) applying the prolongation ${\bf pr} (\pmb{v})$ of the vector field $\pmb{v}$:
\be \label{v}
 \pmb{v} = \xi \partial_u + \eta \partial_v + \phi \partial_\theta + B_1 \partial_{k_1} + B_2 \partial_{k_2} \ ,
\ee
where $\xi$, $\eta$ and $\phi$ are functions of $u, v, \theta, k_1$ and $k_2$. The prolongation ${\bf pr} (\pmb{v})$ is computed in the standard way:
\[
{\bf pr} (\pmb{v}) = \pmb{v} +  \phi^{u} \partial_{\theta_{,u}} + \phi^{v} \partial_{\theta_{,v}} + \phi^{uu} \partial_{\theta_{,uu}} + \phi^{uv} \partial_{\theta_{,uv}}  + B_1^u \partial_{k_{1,u}} + B_2^v \partial_{k_{2,v}}  , 
\]  
where other terms are omitted because $\theta_{_,vv}$, $k_{1,u}$, $k_{2,v}$ and higher derivatives can be eliminated using \rf{gmc}. W recall that 
\be \ba{l}
   \phi^{u} := D_u (\Phi - \xi \theta_{,u} - \eta \theta_{,v}) + \xi \theta_{,uu} + \eta \theta_{,uv} \ ,  \\[1ex]
     B_1^u := D_u (B_1 - \xi k_{1,u} - \eta k_{1,v}) + \xi k_{1,u u} + k_{1,uv} ,  \\[1ex] 
  \phi^{uv} := D_{u v} (\Phi - \xi \theta_{,u} - \eta \theta_{,v}) + \xi \theta_{,uuv} + \eta \theta_{,uvv} \ , \\[1ex]
 \ea \ee
and analogous formulae for $\phi^{u}$, $\phi^{uu}$ and $B_2^v$, see \cite{O}. Finally, $D_u$ denotes a total derivative with respect to $u$. 
 
Standard symmetry procedures (see, for example \cite{O}) yield the system of the determining equations
\be \ba{l}
\label{determ}
\xi = \xi (u),\quad \eta = \eta (v),\quad \phi = \phi (\theta)\\[1ex]
 B_1 = B_1 (v, k_1 , \theta),\quad B_2 = B_2 (u, k_2, \theta), \\[1ex]
 \xi_{,uu} = \eta_{,vv} = \phi_{,\theta \theta} = 0,\quad \xi_{,u} = \eta_{,v}, \\[1ex]
 0 = B_{1,\theta} + B_1 - B_2 + (k_1 - k_2) (\phi_{,\theta} - B_{1, k_1}), \\[1ex]
 0 = B_{2,\theta} + B_2 - B_1 + (k_2 - k_1) (\phi_{,\theta} - B_{2, k_2}), \\[1ex]
 0 = 2 \phi - \phi_{,\theta}+ 2 \xi_{,u} + \frac{B_1}{k_1} + \frac{B_2}{k_2} .
\ea \ee 

To obtain these we need to prolong two last equations of (\ref{gmc}). We point out that this pair of equations is invariant with respect to simultaneous  change of variables $u \leftrightarrow v$ and $k_1 \leftrightarrow k_2$. Using ordinarily understood linear independence for differential polynomials, we obtain  that $B_1 = B_1 (v, k_1 , \theta), B_2 = B_2 (u, k_2, \theta)$, and $\xi = \xi (u), \eta = \eta (v)$. Here also we find $\phi = \phi (\theta)$.   To get the last line, and the condition $\xi_{u} = \eta_{,v}$, together with $\xi_{,uu} = 0 = \eta_{,vv}$, we need to use the first equation of (\ref{gmc}) and substitute it into its prolonged version (eliminating, for instance, $\theta_{,vv}$).

Solving equations \rf{determ} we get
\be \ba{l} 
\xi = \xi (u),\quad \xi_{,uu} = 0 \ \ \rightarrow \ \ \xi = c_1 + c_0 u, \\[1ex]
 \eta = \eta (v),\quad \eta_{,vv} = 0 \ \ \rightarrow \ \ \eta = c_2 + \tilde{c}_0 v, \\[1ex]
 \xi_{,u} = \eta_{,v}  \ \ \rightarrow \ \ \tilde{c}_0 = c_0,\quad \phi_{,\theta \theta} = 0  \ \ \rightarrow \ \ \phi = c_3 + c_5 \theta, \\[1ex]
 B_1 = B_1 (v, k_1 , \theta),\quad B_2 = B_2 (u, k_2, \theta) ,
\ea \ee
where $c_0, c_1, c_2, c_3$ and $c_5$ are constant. 
Note that conditions $\xi_{,uu} = 0 = \eta_{,vv}$ follow from  $\xi_{,u} = \eta_{,v}$ since each of $\xi, \eta$ is a function of one variable. 
The last line of \rf{determ} implies that
\be
B_1 = k_1 A_1 (\theta, v) , \quad B_2 = k_2 A_2 (\theta, u) . 
\ee
and substituting these into the preceding two lines of \rf{determ}, and once again into the last line, we obtain
\be
  A_2 = A_1 = c_4 \ ,  \quad c_5 = 0 \ ,
\ee
where $c_4$ is constant. Finally, we arrive at the following result.

\begin{Th}   \label{Th-sym-3}
The Gauss-Codazzi equations for isothermic surfaces (\ref{gmc}) have 4-parameter group of Lie point symmetries given by 
\be \ba{l} \label{sym-gen}
\xi = c_1 + c_0 u,\quad \eta = c_2 + c_0 v, \quad  \phi = c_3  \\[1ex]
 B_1 = - (c_0 + c_3) k_1, \quad B_2 = - (c_0 + c_3)  k_2 \ ,  
\ea \ee
with $c_j$ constant for $j=1,2,3,4$.
\end{Th}

Thus we obtained four vector fields
\be \ba{l}  \label{4sym}
 v_1 =  \partial_u,\\[1ex]
 v_2 =  \partial_v, \\[1ex]
 v_3 = \partial_\theta - k_1 \partial_{k_1} - k_2 \partial_{k_2} , \\[1ex]
 v_0 =  u \partial_u + v \partial_v  - k_1 \partial_{k_1} - k_2 \partial_{k_2} ,  
\ea \ee
which have simple geometric interpretation: $v_1, v_2$ are translations in $u, v$ coordinates, respectively; $v_3$ is a scaling in coordinates $u, v$  performed simultaneously with  translation in variable $\theta$; and, finally, $v_0$ is a scaling in variables $u, v, k_1$ and $k_2$.

\section{Lie point symmetries of non-parametric linear problems for isothermic surfaces}

Let us consider a nonlinear system, denoted by $\Delta = 0$, which arises as compatibility conditions 
\be \label{nonlin}
   U_2,_1 - U_1,_2 +[U_1, U_2] = 0 
\ee
for a system of linear equations
\be  \label{problin}
  \Psi,_1 = U_1 \Psi \ , \quad \Psi,_2 = U_2 \Psi \ ,
\ee 
where $U_1$ and $U_2$ depend on $x^1 \equiv u$ and $x^2 \equiv v$ also through dependent variables $\varphi^\alpha$ ($\alpha = 1,2,\ldots$), like $\theta$, $k_1$ and $k_2$. Then, we consider transformations which are point transformations (defined by   $\pmb{v}$) with respect to dependent and independent variables  and gauge transformations with repect to $\Psi$ (i.e.,  $\tilde \Psi = G \Psi$).  

Following \cite{Ci-isolate} and \cite{CGS2}, we can apply infinitesimal version of the gauge transformation: $G = I + \ep M + \ldots$, where $\ep$ is the corresponding group parameter. Thus we have to use 
the prolongation ${\bf pr} (\pmb{v})$ of the vector field $\pmb{v}$ (e.g., \rf{v}) with additional, extra-term $M \Psi \partial_\Psi$, where $M$ is a matrix depending on independent and dependent variables of the considered nonlinear system \rf{nonlin}, and $\partial_\Psi$ denotes the partial derivative with respect to entries of the wave function. 

\begin{prop}[\cite{Ci-isolate}]
We consider a transformation of \rf{problin} which is a gauge transformation with respect to $\Psi$ and and a Lie point transformation with respect to variables $x^\mu$ ($\mu =1,2$) and $\varphi^\alpha$ ($\alpha =1,2\ldots$). This transformation preserves the linear problem \rf{problin} if and only if  
\be  \label{DM}
D_k (M) = \left( [U_k, M] + {\bf pr} (\pmb{v}) (U_k) + D_k (\xi^1) U_1 + D_k (\xi^2) U_2 \right) \left|_{\Delta = 0}  \right. . 
\ee
where $D_k$ ($k=1,2$) is the total derivative with respect to $x^k$, $M$ is a generator of the gauge transformation and, finally, $\pmb{v} = \xi^k \partial_k + \Phi^\alpha \partial_{\varphi^\alpha}$ (where summation over repeating indices is assumed) is a generator of the Lie point transformation. 
\end{prop}

We point out that the most interesting case is when \rf{DM} is not solvable for a vector field $\pmb{v}$ which is a symmetry of \rf{nonlin}.   

\begin{cor}
If ${\bf pr} (\pmb{v}) (\Delta)|_{\Delta=0} = 0$ and \rf{DM} has no solution for this $\pmb{v}$, then $\pmb{v}$ introduces a non-removable parameter into the linear problem \rf{problin}. 
\end{cor}

One has to remember that this does not prove that the considered problem is integrable.  The integrability has to be shown separately (e.g., by constructing the Darboux transformation or the recursion operator).

\subsection{Lie point symmetries of $\mathfrak so(3)$-valued linear problem}

Fundamental forms \rf{forms} apparently define a geometric problem in $\mathbb E^3$. The corresponding Gauss-Weingarten equations (which define a kinematics of the moving frame) read
\be
    \Phi,_u = \check{U}_u \Phi , \qquad \Phi,_v = \check{U}_v \Phi , 
\ee
where $\Phi \in SO(3)$ and 
\be \ba{l}    \label{Lax3}
\check{U}_u = \left( \ba{ccc} 0 & - \theta_{,v} & - k_2 e^\theta \\ \theta_{,v} & 0 & 0  \\ k_2 e^\theta & 0 & 0 \ea \right)  , \\[6ex] 
\check{U}_v = \left( \ba{ccc} 0 & \theta_{,u} & 0 \\ - \theta_{,u} & 0 & - k_1 e^\theta  \\ 0 & k_1 e^\theta & 0   \ea \right) .
\ea \ee
All symmetries \rf{4sym} leave the Lax pair \rf{Lax3} invariant. Therefore, even without using equations \rf{DM}, we have the following conclusion.  

\begin{cor}
The spectral parameter cannot be inserted into the $\mathfrak so(3)$ linear problem by Lie point symmetries.  
\end{cor}

\subsection{$\mathfrak so(4,1)$-valued non-parametric linear problem} 

Then, motivated by \cite{Ci-DGA} and \cite{CGS1}, we may consider another linear problem of the form \rf{problin}, in a larger space: 
\be \ba{l} \label{problin-5}
\Psi,_u  = \left( \ba{ccccc} 0 & - \theta_{,v} & - k_2 e^\theta &  \sinh \theta  & \cosh\theta \\ \ \theta_{,v} & \quad 0 & \quad 0 & 0 & 0 \\ \ k_2 e^\theta & \quad 0 & \quad 0 & 0 & 0 \\ - \sinh \theta  & \quad 0 & \quad 0 & 0 & 0 \\ \quad \cosh \theta & \quad 0 & \quad 0 & 0 & 0 \ea \right) \Psi , \\[8ex]
\Psi,_v  = \left( \ba{ccccc} \ 0 & \quad \theta_{,u} & \quad 0 &  0 & 0   \\ \ - \theta_{,u} & \quad 0 & - k_1 e^\theta & \cosh\theta  & \sinh\theta \\ \ 0 & \quad k_1 e^\theta & \quad 0 & 0 & 0 \\  \ 0  &  -\cosh\theta & \quad 0 & 0 & 0 \\ \ 0 & \quad \sinh\theta & \quad 0 & 0 & 0 \ea \right)  \Psi , 
\ea \ee
with compatibility conditions \rf{nonlin} taking exactly the form (\ref{gmc}).

We are going to apply the algorithm presented in \cite{CGS2}, based on simple observation that if  Lie symmetries of the nonlinear system (here (\ref{gmc})) and its linear problem  differ by some vector field, we can use this vector field to introduce a parameter which cannot be gauged out (non-removable parameter). To compute symmetries of the linear problem we use \rf{DM}. 

However, calculations are much convenient if we use an isomorphic form of the linear problem \rf{problin-5} with values in the Clifford algebra $\Cl (4,1)$ (more precisely: in the spin$(4,1)$ Lie algebra). In the next section we will discuss this case in more detail.

\subsection{$\Cl (4,1)$-valued non-parametric linear problem}

Motivated by \cite{Ci-TMP} (see also \cite{CGS1}), we consider the linear problem \rf{problin}, where
\be
\label{lpc}
\ba{l}  U_1 = \frac{1}{2} {\bf{e}}_1 \bigg( \alpha_{2} {\bf{e}}_2 + \alpha_{3} {\bf{e}}_3 + \alpha_{4} {\bf{e}}_4 + \alpha_{5} {\bf{e}}_5 \bigg),   \\[3ex]
   U_2 = \frac{1}{2} {\bf{e}}_2 \bigg( \beta_{1} {\bf{e}}_1 + \beta_{3} {\bf{e}}_3 + \beta_{4} {\bf{e}}_4 + \beta_{5} {\bf{e}}_5 \bigg), 
\ea \ee
and $\e_1,\ldots,\e_5$ are generators of the Clifford algebra $\Cl (4,1)$, i.e., 
\[
\e_k \e_j + \e_j \e_k = 0 \ \ (\text{for} \ j\neq k), \quad \e_1^2 = \e_2^2 = \e_3^2 = \e_4^2 = - \e_5^2 = 1 . 
\]
 Compatibility conditions \rf{nonlin}  
yield the following system of differential equations for coefficients $\alpha_{j}, \beta_{j}$:
\be
\label{cl}
\ba{l}  0 = \alpha_{2,v} + \beta_{1,u} - \alpha_{3} \beta_{3} - \alpha_{4} \beta_{4} + \alpha_{5} \beta_{5}, \\[1ex]
 0 = \alpha_{3,v} + \alpha_{2} \beta_{3},  \\[1ex]
  0 = \beta_{3,u} +  \beta_{1} \alpha_{3},    \\[1ex]
0 = \alpha_{4,v} + \alpha_{2} \beta_{4},  \\[1ex]
 0 = \beta_{4,u} +  \beta_{1} \alpha_{4},   \\[1ex]
 0 = \alpha_{5,v} + \alpha_{2} \beta_{5},  \\[1ex]
 0 = \beta_{5,u} +  \beta_{1} \alpha_{5}. 
\ea \ee

The full system \rf{cl} is not equivalent to \rf{gmc}. In order to get an equivalent system we have to add some constraints. 

\begin{rem}
It can be easily seen that the substitution
\be \ba{l}
\alpha_{2} = - \theta_{,v},\quad \beta_{1} = - \theta_{,u},\quad \alpha_{3} = - k_2 e^\theta,\quad \beta_{3} = - k_1 e^\theta, \\[1ex]
\alpha_{4} = \alpha_{5} = \beta_{4} = \beta_{5} = 0 \ ,
\ea \ee
leads us back to the Gauss-Codazzi equations \rf{gmc}. However, this reduction is too restrictive because the linear problem becomes equivalent to the previously considered $\mathfrak{so} (3)$ linear problem (without a possibility to introduce a spectral parameter by Lie point symmetries). 
\end{rem}

From the last four equations of \rf{cl} we derive:
\be \ba{l}  \label{hyp}
(\alpha_{4}^2 - \alpha_{5}^2)_{,v} = 2 \alpha_{2} (\alpha_{5} \beta_{5} - \alpha_{4} \beta_{4}) , \\[1ex]
(\beta_{4}^2 - \beta_{5}^2)_{,u} = 2 \beta_{1} (\alpha_{5} \beta_{5} - \alpha_{4} \beta_{4}) .
\ea \ee
This result, together with the form of the first equation of \rf{cl}, is a motivation for introducing the following quadratic constraint: 
\be  \label{quad}
  \alpha_{5} \beta_{5} - \alpha_{4} \beta_{4} = 0 .
\ee
Then, automatically,
\be \label{norma}
  \alpha_{5}^2 - \alpha_{4}^2 = f (u) , \quad \beta_{4}^2 - \beta_{5}^2 = g (v) ,
\ee
where $f$ and $g$ are functions of one variable. The choice $f(u) = g (v) =1$ will lead us to the Lax pair first presented in \cite{CGS1}. Indeed, then from \rf{norma} and \rf{quad} we obtain
\be
\alpha_4 = \sinh \theta, \quad \alpha_5 = \cosh \theta, \quad \beta_4 = \cosh \theta, \quad \beta_5 = \sinh \theta,
\ee
what was considered earlier \cite{CGS1}, and here it will be the only case of  our interest.  This is almost unique solution of the system \rf{quad} and \rf{norma} (for $f(u) = g (v) =1$). The only arbitrary assumptions are $\alpha_5 > 0$ and $\beta_4 > 0$. 
We derive from \rf{cl}: $\alpha_{2} = - \theta_{,v}$ and $\beta_{1} = - \theta_{,u}$, and denote $\alpha_{3} = - k_2 e^\theta$ and $\beta_{3} = - k_1 e^\theta$. Then 
\be \ba{l}  \label{problin-theta}
 U_1 = \frac{1}{2} {\bf{e}}_1 \bigg( - \theta_{,v} {\bf{e}}_2 - k_2 e^\theta {\bf{e}}_3 +  {\bf{e}}_4 \sinh \theta +  {\bf{e}}_5 \cosh \theta  \bigg), \\[3ex]
 U_2 = \frac{1}{2} {\bf{e}}_2 \bigg( - \theta_{,u} {\bf{e}}_1 - k_1 e^\theta {\bf{e}}_3 +  {\bf{e}}_4 \cosh \theta +  {\bf{e}}_5 \sinh \theta  \bigg).
\ea \ee
The linear problem \rf{problin-theta} is equivalent to \rf{problin-5} because the Lie algebra $\mathfrak so$(4,1) is isomorphic to the Lie algebra of the Lie group Spin(4,1). This group, in turn, is a double covering of $SO(4.1)$.

\subsection{Lie point symmetries of $\Cl (4,1)$-valued linear problem}

The Lie point symmetries of Gauss-Codazzi equations are given by the following vector field
\be \label{v-sym-3}
 \pmb{v} = (c_1 + c_0 u) \partial_u + (c_2 + c_0 v) \partial_v + c_3 \partial_\theta - (c_0 + c_3) (k_1 \partial_{k_1} + k_2 \partial_{k_2})  ,
\ee
see \rf{sym-gen}. We are going to check which transformations of the form \rf{v-sym-3} are Lie point symmetries of the non-parametric linear problem \rf{problin-theta}.  The system \rf{DM} assumes the form:
\be \ba{l}  \label{DM-3}
M,_\theta \theta,_u + M,_{k_1} k_1,_u + M,_{k_2} k_2,_u + M,_u = [U_1, M] + {\bf pr} (\pmb{v}) (U_1) + c_3 U_1 \ , \\[1ex]
M,_\theta \theta,_v + M,_{k_1} k_1,_v + M,_{k_2} k_2,_v + M,_v = [U_2, M] +{\bf pr} (\pmb{v}) (U_2) + c_3 U_2 . 
\ea \ee
The last parts of these equations can be computed as
\[ \ba{l}
{\bf pr} (\pmb{v}) (U_1) + c_0 U_1 = \frac{1}{2} \e_1 \left( (c_0 \sinh\theta + c_3 \cosh\theta) \e_4 + (c_3 \sinh\theta + c_0 \cosh\theta) \e_5 \right) , \\[1ex]
{\bf pr} (\pmb{v}) (U_2) + c_0 U_2 = \frac{1}{2} \e_2 \left( (c_3 \sinh\theta + c_0 \cosh\theta) \e_4 + (c_0 \sinh\theta + c_3 \cosh\theta) \e_5 \right) ,
\ea \]
where we took into account that 
the first prolongation of $\pmb{v}$ is given by: 
\be
 {\bf pr}^{1} (\pmb{v}) = \pmb{v} + \Phi^u \partial_{\theta,_u}  + \Phi^v \partial_{\theta,_v}  = \pmb{v}  - c_0 \theta,_u \partial_{\theta,_u}  - c_0 \theta,_v \partial_{\theta,_v} .
\ee
Equating to zero coefficients by (independent) derivatives in equations \rf{DM-3} we obtain:
\be
M = M (u, v) \ , \quad [M, \e_1 \e_2] = 0 .
\ee
The, considering terms by $k_1$ and $k_2$ we get
\be
  [M, \e_1 \e_3] = 0 \ , \quad [M, \e_2 \e_3] = 0 .
\ee
Therefore, $M$  (as a bivector) has to be of the form:
\be
  M = c_4 \e_4 \e_5 ,
\ee
where $c_4$ is constant (which follows from the terms by $\e_4 \e_5$). Finally, the system \rf{DM-3} splits into 4 equations (coefficients by $\e_1\e_4$, $\e_1 \e_5$, $\e_2\e_4$ and $\e_2 \e_5$, respectively), but only two of them are different:  
\be \ba{l}
0 =  c_4 \cosh \theta  + \frac{1}{2} \left(  c_3 \cosh \theta + c_0 \sinh\theta \right) , \\[1ex]
0 = c_4 \sinh\theta + \frac{1}{2} \left( c_3 \sinh \theta + c_0 \cosh \theta \right) .
\ea \ee
Hence, finally,
\be
  c_0 = 0 , \quad c_4 = - \frac{1}{2}  c_3 .
\ee
Therefore, a non-removable parameter can be introduced by the vector field $v_0$. Thus we obtain the linear problem
\be \ba{l}  \label{problin-theta-param}
 U_1 = \frac{1}{2} {\bf{e}}_1 \bigg( - \theta_{,v} {\bf{e}}_2 - k_2 e^\theta {\bf{e}}_3 + \lambda {\bf{e}}_4 \sinh \theta  + \lambda {\bf{e}}_5 \cosh \theta  \bigg), \\[3ex]
 U_2 = \frac{1}{2} {\bf{e}}_2 \bigg( - \theta_{,u} {\bf{e}}_1 - k_1 e^\theta {\bf{e}}_3 + \lambda {\bf{e}}_4 \cosh \theta  + \lambda {\bf{e}}_5 \sinh \theta  \bigg),
\ea \ee
where $\lambda = e^\ep$ and $\ep$ is a parameter of one parameter group corresponding to the vector field $v_0$.

We recall that the Lax pair \rf{problin-theta-param} was first derived in \cite{CGS1}  from the system of linear equations with a parameter known to Darboux and Bianchi  \cite{Bi}, by a change of variables and a reparameterization of the parameter. The theory of isothermic surfaces can be extended on immersions in multidimensional Euclidean spaces (see, for instance, \cite{Bu} and referencces cited therein).

\section{Isothermic surfaces in $\E^n$}

The approach presented in previous sections can be extended on the multidimensional case. We consider the following linear problem:
\be
\label{problin-Rn}
\ba{l}  U_1 = \frac{1}{2} {\bf{e}}_1 \left( - \theta,_v {\bf{e}}_2 + \alpha_{3} {\bf{e}}_3 + \ldots + \alpha_{n} {\bf{e}}_n +  {\bf{e}}_{n+1}\sinh\theta + \e_{n+2}\cosh\theta  \right) ,   \\[2ex]
   U_2 = \frac{1}{2} {\bf{e}}_2 \left( - \theta,_u  {\bf{e}}_1 + \beta_{3} {\bf{e}}_3 + \ldots + \beta_{n} {\bf{e}}_n + {\bf{e}}_{n+1}\cosh\theta + \e_{n+2}\sinh\theta \right), 
\ea \ee
where $\e_1,\ldots,\e_{n+1}$ are generators of the Clifford algebra $\Cl (n+1,1)$, i.e., 
\[
\e_k \e_j + \e_j \e_k = 0 \ \ (\text{for} \ j\neq k), \quad \e_k^2 = 1  \ \ (\text{for} \ 1\leqslant  k \leqslant n+1), \quad \e_{n+2}^2 = -1 . 
\]
 Compatibility conditions \rf{nonlin}  
yield the following system of differential equations for coefficients $\alpha_{j}, \beta_{j}$ and $\theta$:
\be
\label{cc-Rn}
\ba{l}  
\theta,_{uu} + \theta,_{vv} + \alpha_3 \beta_3 + \ldots + \alpha_n \beta_n  = 0  \\[1ex]
\alpha_j,_v = \theta,_v  \beta_{j} \qquad \ (j =3,\ldots, n) ,  \\[1ex]
\beta_j,_u = \theta,_u \alpha_j  \qquad \ (j =3,\ldots, n) .
\ea \ee

\subsection{Lie point symmetries of Gauss-Codazzi equations}

Now we are going to compute the symmetries of the system (\ref{cl}) with and without the constraint \rf{quad}. To do this we need the following prolongation of a general  vector field $\pmb{v}$:
\be \ba{l}
\lefteqn{ {\bf pr} (\pmb{v}) = \xi \partial_u + \eta \partial_v + \Phi \partial_\theta +  A^{j} \partial_{\alpha_{j}} + B^{j} \partial_{\beta_{j}}  + \Phi^u \partial_{\theta,_u} + \Phi^v \partial_{\theta,_v} {}}  \\[1ex]
 {} \quad + A^{ju} \partial_{\alpha_{j,u}} + B^{ju} \partial_{\beta_{j,u}} + A^{jv} \partial_{\alpha_{j,v}} + B^{j v} \partial_{\beta_{j,v}} + \Phi^{uu} \partial_{\theta,_{uu}} + \Phi^{vv} \partial_{\theta,_{vv}} ,\;
\ea \ee
where all coefficients of the vector field $\pmb{v}$ depend on $u,v, \theta$, $\alpha_{j}$ and $\beta_{j}$, with $j = 3,\ldots,n$. Moreover, 
\be  \ba{l}
\Phi^u = D_u \Phi - (D_u \xi) \theta,_u -  (D_u \eta) \theta,_v , \\[2ex]
\Phi^v = D_v \Phi - (D_v \xi) \theta,_u -  (D_v \eta) \theta,_v , \\[2ex]
A^{jv} = D_v A^j - \alpha_j,_u D_v \xi - \alpha_j,_v D_v \eta , \qquad \ (j =3,\ldots, n) ,  \\[2ex]
B^{ju} = D_u B^j - \beta_j,_u D_u \xi - \beta_j,_v D_u \eta , \qquad \ (j =3,\ldots, n) , \\[2ex]
\Phi^{uu} = D_{uu} \Phi - (D_{uu} \xi) \theta,_u - 2 (D_u \xi) \theta,_{uu} - (D_{uu} \eta) \theta,_v - 2 (D_u \eta) \theta,_{uv} , \\[2ex]
\Phi^{vv} = D_{vv} \Phi - (D_{vv} \xi) \theta,_u - 2 (D_v \xi) \theta,_{uv} - (D_{vv} \eta) \theta,_v - 2 (D_v \eta) \theta,_{vv} .  
\ea \ee
Applying the prolonged vector field ${\bf pr} (\pmb{v})$ to the nonlinear system \rf{cc-Rn} we obtain
\be \ba{l}   \label{symm-Rn}
\Phi^{uu} + \Phi^{vv} + A^k \beta_k + \alpha_k B^k = 0 \ , \\[2ex]
A^{jv} = \Phi^v \beta_j + \theta,_v B^j \qquad \ (j =3,\ldots, n) ,  \\[2ex]
B^{ju} = \Phi^u \alpha_j + \theta,_u A^j \qquad \ (j =3,\ldots, n) , 
\ea \ee
where the Einstein summation convention is used (in the first equation). 

The standard computation of Lie symmetries proceeds as follows. First, we consider coefficients by $\alpha_j,_u$ and $\beta_j,_v$ in the second and third set of equations \rf{symm-Rn}, respectively. The result is
\be \label{zale}
   \xi = \xi (u) , \qquad \eta = \eta (v) \ .
\ee
Then, we equate to zero coefficients by $\alpha_j,_{uu}$ and $\beta_j,_{vv}$. Hence, $\Phi,_{\alpha_j} = 0$ and $\Phi,_{\beta_j} = 0$. Therefore, 
\be \label{fid}
    \Phi = \Phi (u, v, \theta) . 
\ee
Using this information, we consider coefficients by $\beta_j,_v$ and $\alpha_j,_u$   in the second and third set of equations \rf{symm-Rn}, respectively. Hence, 
\be
A^j,_{\beta_k} = 0 , \qquad B^j,_{\alpha_k} = 0 ,
\ee 
for any $j, k$. Substituting $\theta,_{uu} + \theta,_{vv}$ from \rf{cc-Rn} and splitting the first equation of \rf{symm-Rn} into two equations (the second one contains first derivatives of $\theta$) we rewrite the system \rf{symm-Rn} as:
\be \label{deter} \ba{l}
\Phi,_{uu} + \Phi,_{vv} - \Phi_\theta (\alpha_k \beta_k) - 2 \dot\eta \theta,_{vv} - 2 \xi' \theta,_{uu} +   A^k \beta_k + \alpha_k B^k = 0 , \\[2ex]
\Phi,_{\theta\theta} (\theta^2_{,u} + \theta^2_{,v}) + 2 \Phi,_{\theta u} \theta,_u + 2 \Phi,_{\theta v} \theta,_v - \ddot\eta \theta,_v - \xi'' \theta,_u = 0 , \\[2ex]
A^j_{,v} + A^j_{,_\theta} \theta,_v + A^j_{\alpha_k} \theta,_v \beta_k = (\Phi,_v + \Phi,_\theta \theta,_v) \beta_j + B^j \theta,_v , \\[2ex]
B^j_{,u}+ B^j_{,_\theta} \theta,_u + B^j_{\beta_k} \theta,_u \alpha_k = (\Phi,_u + \Phi,_\theta \theta,_u) \alpha_j + A^j \theta,_u ,
\ea \ee
where, again, we assume summation over repeating index $k$ (from 3 to $n$).
Then, we use (\ref{cc-Rn}) for eliminating $\theta_{,vv}$ from (\ref{deter}). The  terms by $\theta_{,uu}$ yield 
\be
\eta,_v = \xi,_u  = c_0,
\ee
where we took into account (\ref{zale}).

This in turn implies that $\eta$ and $\xi$ are linear functions of its variables, i.e.
\be
\xi (u) = c_0 u + c_1, \qquad \eta (v) = c_0 v + c_2 .
\ee
Using (\ref{fid}) and the term by squares of $\theta$ derivatives in the second equation of (\ref{deter}) we obtain 
\be
\Phi = c_4 (u,v) \theta (u,v) + c_3 (u,v) .
\ee
Looking again into (\ref{deter}) we see that $\Phi_{,\theta} = \const$, and the last two equations in there tell us (through terms by $\alpha_j, \beta_j$) that
\be
\Phi = c_3 = \const.
\ee
Then, from the last two equation of \rf{deter} we get 
\be
   A^j_{,v} = 0 \ , \quad B^j_{,u} = 0 , 
\ee
and, finally, we are left with
\be \ba{l}
0 = 2 c_0 \alpha_k \beta_k + \alpha_k B^k + A^k \beta_k ,  \\[1ex]
B^j = A^j_{\alpha_k} \beta_k + A^j_{,\theta} \ , \\[1ex]
A^j = B^j_{\beta_k} \alpha_k + B^j_{,\theta} 
\ea \ee
It is not difficult to check that the most general solution for this system is
\be
A^j = a^{jk} \alpha_k - c_0 \alpha_j, \qquad B^j = a^{jk} \beta_k - c_0 \beta_j ,
\ee
with $a^{jk}$ forming an antisymmetric constant matrix.

\begin{Th}  \label{Th-sym-n}
The system of determining equations \rf{symm-Rn} has general solution of the form 
\be \ba{l}
\label{th1}
\xi = c_0 u + c_1 \ , \qquad \eta = c_0 v + c_2 \ , \qquad \Phi = c_3 \ , \\[1ex]
A^j = a^{jk} \alpha_k - c_0 \alpha_j \ , \qquad B^j = a^{jk} \beta_k - c_0 \beta_j ,
\ea \ee
where $c_0$, $c_1$, $c_2$, $c_3$ and $a^{jk}$ ($k=3,\ldots,n$, $j\neq k$) are constants such that
\[
   a^{jk} = - a^{kj} .
\]
\end{Th}

\no Theorem~\ref{Th-sym-3} is a special case of Theorem~\ref{Th-sym-n} but  parameterizations of dependent variables is slightly different in both cases.

\subsection{Lie point symmetries of the $\Cl (n+1,1)$-valued non-para\-metric linear problem}

Computing Lie point symmetries of the linear problem \rf{problin-Rn} we will need   coefficients $\Phi^u$ and $\Phi^v$ of the first prolongation of $\pmb{v}$.  We have
\be   \label{prol-Phi}
\ba{l}
\Phi^u = D_u (\Phi - \xi \theta_{,u} - \eta \theta_{,v}) + \xi \theta_{,uu} + \eta \theta_{uv} = - c_0 \theta_{,u} \\ [1ex]
\Phi^v = D_v (\Phi - \xi \theta_{,u} - \eta \theta_{,v}) + \xi \theta_{,uv} + \eta \theta_{vv} = - c_0 \theta_{,v},
\ea
\ee
where we substituted solutions for coefficients \label{th1} of the vector field $\pmb{v}$. On the other hand, the system (\ref{DM}) becomes 
\be
\label{DM-n}
\ba{l}
M_{,u} + M_{,\theta} \theta_{,u} + M_{,\alpha_j} \alpha_{j,u} + M_{,\beta_j} \theta,_u \alpha_j  = [U_1, M] + {\bf pr} {\pmb v} (U_1) + c_0  U_1, \\ [1ex]
M_{,v} + M_{,\theta} \theta_{,v} + M_{,\alpha_j} \theta,_v \beta_j  + M_{,\beta_j} \beta_{j,v} = [U_2, M] + {\bf pr} {\pmb v} (U_2) + c_0  U_2 .
\ea
\ee
Using \rf{prol-Phi} and \rf{th1}, we compute
\be
\label{prole}
\ba{l}
{\bf pr} {\pmb v} (U_1) + c_0 U_1 = \frac{1}{2} \e_1 \e_j a^{jk}  \alpha_k \\[1ex] \qquad 
+ \frac{1}{2}  \e_1 \left( (c_0 \e_{n+1} + c_3 \e_{n+2}) \cosh\theta +   (c_3 \e_{n+1} + c_0 \e_{n+2} ) \sinh\theta  \right) ,    \\ [2ex]
{\bf pr} {\pmb v} (U_2) + c_0 U_2 = \frac{1}{2} \e_2 \e_j a^{jk}  \beta_k \\[1ex] \qquad 
+ \frac{1}{2}  \e_2 \left(  (c_3 \e_{n+1} + c_0 \e_{n+2}) \cosh\theta +   (c_0 \e_{n+1} + c_3 \e_{n+2} ) \sinh\theta \right)  .
\ea
\ee
Commutators in \rf{DM-n} do not contain derivatives except, respectively, $\theta,_v$ and $\theta,_u$. Therefore, considering coefficients by derivstives in \rf{DM-n}, we have
\be  \label{cond}
 M = M(u,v) , \qquad [M, \e_1 \e_2] = 0 .
\ee
Then, considering coefficients by $\cosh\theta$, $\sinh\theta$, $\alpha_k$, $\beta_k$ and terms independent of these variables, we reduce the system \rf{DM-n} to 
\be \ba{l}  \label{DM-sol}
[M,\e_1 \e_{n+1}] = c_0 \e_1 \e_{n+1} + c_3 \e_1 \e_{n+2} , \\[1ex]
[M,\e_1 \e_{n+2}] = c_3 \e_1 \e_{n+1} + c_0 \e_1 \e_{n+2} , \\[1ex]
[M,\e_2 \e_{n+2}] = c_3 \e_2 \e_{n+1} + c_0 \e_2 \e_{n+2} , \\[1ex]
[M,\e_2 \e_{n+1}] = c_0 \e_2 \e_{n+1} + c_3 \e_2 \e_{n+2} , \\[1ex]
[M, \e_1 \e_k] = \e_1 \e_j a^{jk} ,  \\[1ex]
[M, \e_2 \e_k] = \e_2 \e_j a^{jk}  ,  \\[1ex]
M,_u = 0 = M,_v ,
\ea \ee
where we assume summation (from 3 to $n$) over repeating index $j$. 
$M$ is a bivector (a linear combination of $\e_\mu \e_\nu$ for $\mu, \nu = 1,2,\ldots,n+2$). Thus
\be
  M =    \frac{1}{2} M^{\mu \nu} \e_\mu \e_\nu ,  \qquad (M^{\mu\nu} = - M^{\nu\mu}) , 
\ee
or, in other words, $M = M^{\mu \nu} \e_\mu \e_\nu$ (where $\mu < \nu$).  
The second condition \rf{cond} means that 
\be
   M^{1\nu} = M^{2 \nu} = 0 \qquad (\text{for} \ \ \nu = 3,\ldots,n+2) . 
\ee
Each of the first four equations of \rf{DM-sol} yields the same conclusion:
\be  \label{c0=0}
    M^{n+1, n+2} = - \frac{1}{2} c_3 \ , \qquad c_0 = 0 . 
\ee
The next two equations of \rf{DM-sol} can be written as:
\be  \label{next-eq}
 \frac{1}{2} [M^{ij} \e_i \e_j , \e_m \e_k] = \e_m \e_j a^{jk} ,
\ee
$m=1,2$ and \ $3 \leqslant i,j,k \leqslant n$. 
In order to solve these equations we need the following identity:
\be
 \frac{1}{2} [\e_i \e_j , \e_m \e_k] =  \delta_{k j} \e_m \e_i  - \delta_{k i} \e_m \e_j ,
\ee
(where $\delta_{k\nu}$ is Kronecker's delta).  Then \rf{next-eq} becomes
\be
 M^{ik} \e_m \e_i - M^{kj} \e_m \e_j = \e_m \e_j a^{jk} .
\ee
Hence
\be
   M^{jk} = \frac{1}{2} a^{jk}
\ee
Therefore, we obtained a series of constraints on the generator $M$ and one restriction on the vector field $\pmb{v}$, namely $c_0 = 0$. 

\begin{cor} 
The vector field $v_0$, given by
\be
v_0 = u \partial_u + v \partial_v - \alpha_j \partial_{\alpha_j} - \beta_j \partial_{\beta_j} ,
\ee
introduces a non-removable parameter into the linear problem \rf{problin-Rn}. \end{cor}

\no Thus we derived the following Lax pair:
\be
\label{problin-Rn-spect}
\ba{l}  U_1 = \frac{1}{2} {\bf{e}}_1 \left( - \theta,_v {\bf{e}}_2 + \alpha_j \e_j  +  \lambda {\bf{e}}_{n+1}\sinh\theta + \lambda \e_{n+2}\cosh\theta  \right) ,   \\[2ex]
   U_2 = \frac{1}{2} {\bf{e}}_2 \left( - \theta,_u  {\bf{e}}_1 + \beta_j \e_j + \lambda {\bf{e}}_{n+1}\cosh\theta + \lambda \e_{n+2}\sinh\theta \right), 
\ea \ee
where summation over $j$ (from $3$ to $n$) is assumed. The Darboux transformation based on this spectral problem was derived in \cite{BC}.

\section{Conclusions}

We studied symmetries of the nonlinear system of Gauss-Codazzi equations for isothermic immersions in $\mathbb E^n$ and symmetries of related non-parametric linear problems. In the case $n=3$ the Lie algebra of point symmetries is 4-dimensional. All these transformations are also symmetries of the associated non-parametric $\mathfrak so (3)$-valued linear problem. Hence, all parameters are removable. Extending the matrix algebra from $\mathfrak so (3)$ to $\mathfrak so (4,1)$ we obtain different situation. The Lie algebra of symmetries of the non-parametric linear problem has the dimension 3. Hence, the fourth vector field introduces the non-removable spectral parameter. We also pointed out that this case can be also studied  by considering more dependent variables subject to appropriate constraints.  Such technique was used in a systematic way to search for integrable reductions of 3-dimensional Gauss-Codazzi equations \cite{KM}. The main new result of our paper is the extension of all these results on the multidimensional case. The problem was formulated in terms of Clifford numbers which simplified greatly all computations. The spectral parameter for isothermic immersions in $\E^n$ was proved to have a group interpretation. 

\ods \ods

\no {\it Acknowledgements.} The research is supported by Polish Ministry of Science and Higher Education. In particular, the second author (A.K.) realizes his research in the framework of the project\ {S/WBiI\'S/3/2016}.

\newpage

\end{document}